\documentclass[a4paper,11pt]{article}
\pdfoutput=1 

\usepackage{jheppub} 
\usepackage{physics}
\usepackage[T1]{fontenc} 
\usepackage{appendix}
\usepackage{braket}

\title{Giant Gravitons, Harish-Chandra integrals, and BPS states in symplectic and orthogonal $\mathcal{N}$= 4 SYM}

\author{Adolfo Holguin,}
\author{Shannon Wang}

\affiliation{Department of Physics, University of California, Santa Barbara, CA 93106, USA}

\emailAdd{adolfoholguin@physics.ucsb.edu}
\emailAdd{shannonwang@physics.ucsb.edu}

\abstract{We find generating functions for half BPS correlators in $\mathcal{N}=4$ SYM theories with gauge groups $Sp(2N)$, $SO(2N+1)$, and $SO(2N)$ by computing the norms of a class of BPS coherent states. These coherent states are built from operators involving Harish-Chandra integrals. Such operators have an interpretation as localized giant gravitons in the bulk of anti-de-Sitter space. This extends the analysis of  \cite{Berenstein:2022srd} to $Sp(2N)$, $SO(2N+1)$, and $SO(2N)$ gauge theories. We show that we may use ordinary Schur functions as a basis for the sector of states with no cross-caps in these theories. This is consistent with the construction of these theories as orientifold projections of an $SU(2N)$ theory.  We make note of some relations between the symmetric functions that appear in the expansion of these coherent states and symplectic Schur functions. We also comment on some connections to Schubert calculus and Gromov-Witten invariants, which suggest that the Harish-Chandra integral may be extended to such problems.}

\begin{document} 
\maketitle
\flushbottom

\section{Introduction}

Recently, there has been a renewed interest in determinant operators in large $N$ holographic gauge theories and their string dual description as giant gravitons \cite{Jiang:2019xdz, Chen:2019gsb, Budzik:2021fyh, Gaiotto:2021xce, Murthy:2022ien}; the dimension of these operators is order $N$, which makes them ideal to probe sub-$AdS$ physics. A natural basis for gauge invariant operators is the Schur functions, which are characters of the unitary and symmetric groups. Combinatorial methods for computing correlation functions in free $\mathcal{N}=4$ SYM were developed in \cite{Corley:2001zk, Corley:2002mj}. More recent works have emphasized the utility of an effective action approach obtained by recasting the determinant operators as fermionic integrals and integrating out the super Yang-Mills fields. In this description, the non-perturbative physics of the problem can be obtained from a saddle point approximation for an effective action in terms of a set of collective fields \cite{Jiang:2019xdz, Budzik:2021fyh}.

A similar prescription for $AdS$ giant gravitons was proposed in \cite{Berenstein:2022srd}, where it was realized that the norms of BPS states are encoded in the expansion of the Harish-Chandra-Itzykson-Zuber (HCIZ) integral, which appears in the evaluation of the norms of a certain class of gauge invariant coherent states:
\begin{equation}
\begin{aligned}
      \mathcal{O}_{\Lambda}(0)&= \int_{SU(N)}dU \exp \left(\Tr\left[ \Lambda U a_Z^\dagger U^\dagger \right] \right).
\end{aligned}
\end{equation}

This sheds light on why the group characters evaluated on the Yang-Mills fields may serve as an orthogonal basis, even though they are only orthogonal with respect to the Haar measure, and gives a different interpretation of the norms of BPS states as the coefficients in the expansion of the HCIZ integral. This technique has the advantage of repackaging the combinatorics of the Schur functions into integrals over the unitary group.

The Harish-Chandra integrals have natural generalizations to the $B$, $C$, $D$ series, $Sp(2N)$ and $SO(M)$. For a choice of simple Lie group $G$, the HCIZ integral has an exact formula in terms of a sum over the saddle points:

\begin{equation}
    \mathcal{H}(x,y) = \int  e^{\langle \text{Ad}_g(x),  y\rangle} dg= c_{\frak{g}}\sum_{w \in W}\frac{\epsilon(w) e^{\langle w(x), y\rangle}}{\Delta_{\frak{g}}(x)\Delta_{\frak{g}}(y)}.
\end{equation}

Each saddle point of the integral corresponds to a Weyl reflection, and the denominators are given by the discriminant of the Lie algebra. These integrals have received less attention than the unitary HCIZ integral, which serves as a single plaquette model in lattice gauge theory.

The bulk of the work on probing finite $N$ physics is limited to field theories with $U(N)$ and $SU(N)$ gauge groups (see \cite{brown2008diagonal, bhattacharyya2008exact, brown2009diagonal, kimura2012correlation}), but more recently, there has been some interest in extending these studies to field theories with $Sp(2N)$, $SO(2N+1)$, or $SO(2N)$ gauge groups \cite{caputa2010spectral, caputa2013basis, caputa2013operators}. There is good reason for this surge of interest: maximally supersymmetric Yang-Mills theory with symplectic and orthogonal groups are dual to type IIB strings on $AdS_5\times \mathbb{RP}^5$ \cite{witten1998baryons}. Depending on the choice of the orientifold projection, the gauge group of the theory is either $Sp(2N)$, $SO(2N+1)$, or $SO(2N)$; $S$-duality relates the spectrum of the $Sp(2N)$ and the $SO(2N+1)$ theories, while the $SO(2N)$ theories are self-dual. The exact matching of the spectrum for the symplectic and orthogonal theories is poorly understood, due to the combinatorial difficulty associated with constructing states of these theories. 

In this paper, we study BPS coherent states of $\mathcal{N}=4$ SYM for special orthogonal and symplectic groups. The norms of such states are given precisely by a Harish-Chandra integral over the corresponding group. By explicitly expanding the integral, we find that these coherent states serve as generating functions for gauge invariant states in the gauge theory, and the corresponding coefficients in the expansion give their norms. In principle, this gives a way of constructing an orthogonal basis of states for these theories from group theoretic data for the corresponding gauge group. We argue that these generating functions are only able to capture information about the "unitary" part of the gauge symmetry, which is to say that operators we find in the expansion match in form to operators in the unitary theory. In section 2, we review the construction of gauge invariant coherent states for the $SU(N)$ theory. In section 3, we generalize this to the symplectic case and argue that the odd special orthogonal case is related to the symplectic case by a rank-level duality that exchanges a Young diagram with its conjugate diagram. We repeat the calculations for the even orthogonal case. In section 4, we discuss other attempts at finding an orthogonal basis for $Sp(2N)$, $SO(2N+1)$, or $SO(2N)$ and how our results can be interpreted in a relevant context. Finally, we conclude with a discussion of a few open questions and future directions of work.

\section{Review of the $U(N)$ case}
\label{sec:SU}
We begin with a brief review of BPS coherent states in $U(N)$. The same analysis may be applied to any free gauge theory with an adjoint scalar field $Z$. We know from \cite{Berenstein:2022srd} that given a na\"\i ve coherent state $F[\Lambda]$ of the form:
\begin{equation}
 \exp( \Tr(\Lambda\cdot a^\dagger_Z))\ket 0,
\end{equation}
where $\Lambda$ is taken to be a diagonal matrix-valued set of parameters and $a^\dagger_z$ is the raising operator for the s-wave of the field $Z$ on $S^3$ in \cite{berenstein2004toy}, we may introduce an auxiliary $U(N)$ group action and average over the group, which allows us to rewrite a gauge invariant coherent state as:
\begin{equation}
\label{BPSoperator}
    F[\Lambda]= \frac{1}{Vol(U(N))}\int dU \exp( \Tr(U\Lambda U^{-1} a^\dagger_Z))\ket 0,
\end{equation}
where $dU$ is the Haar measure. Our normalization factor $Vol(U(N)) = \int dU$; we can set it equal to one for the sake of brevity. We may compute the overlap of $F[\Lambda]$ as defined in Eq. \eqref{BPSoperator} with its adjoint $\bar F[\bar \Lambda]$ by evaluating the HCIZ integral:

\begin{equation}
\label{HCIZintegral}
 \bar F[\bar \Lambda]* F[\Lambda]  = \int d\tilde U\exp\left(\Tr\left( \tilde U ^{-1}\Lambda \tilde U \bar \Lambda^{\prime}\right)\right).
\end{equation}

We see that we have sidestepped most of the Wick contractions of the  matrix operators $\left(a^\dagger\right)^i_j$, which would make $F[\Lambda]$ difficult to compute in the form it takes in Eq. \eqref{BPSoperator}. $F[\Lambda]$ can be evaluated through a character expansion, as described in \cite{morozov2010unitary}:

\begin{equation}
\label{unitaryexpansion}
F[\Lambda]= \sum_R \frac{1}{f_R} \chi_R(\Lambda) \chi_R(a^\dagger_Z)\ket 0
\end{equation}

We may also rewrite Eq. \eqref{HCIZintegral} through a character expansion:

\begin{equation}
\bar F[\bar \Lambda]* F[\Lambda] = \sum_R\frac{1}{f_R}\chi_R\left(\bar{\Lambda}\right)\chi_R\left(\Lambda\right)
\end{equation}

We can then compare the coefficients of the characters from the equation above to what we would obtain from multiplying Eq. \eqref{unitaryexpansion} by its adjoint and find:

\begin{equation}
\bra{0}\chi_R(a)\chi_R(a^\dagger)\ket{0} = f_R
\end{equation}

It becomes obvious that we must compute $f_R$ to evaluate the overlap of $\chi_R(a)$ and $\chi_R(a^\dagger)$. The thing to keep in mind is that the representations $R$ in the coherent state $F[\Lambda]$ correspond to Young diagrams for $U(N)$, which are characterized by the indices $j_1\geq j_2 \geq \dots j_N$, where each index $j_i$ iterates over row $i$. Because these are  characters of the unitary group, they may be rewritten with the Weyl character formula:

\begin{equation}
    \chi_{j_i}(\Lambda) = \frac{\det\left(\lambda_k^{j_i+N-i}\right)}{\Delta(\Lambda)},
\end{equation}
where $\lambda_k$ are the eigenvalues of $\Lambda$ and $\Delta(\Lambda)$ is the Vandermonde determinant of $\Lambda$. Then we may rewrite the HCIZ integral as a product of these expanded characters:

\begin{equation}
 I(\Lambda, \bar \Lambda)= \int d\tilde U\exp\left(\Tr\left( \tilde U ^{-1}\Lambda \tilde U \bar \Lambda^{\prime}\right)\right) = \Omega \frac{\det\left(\exp(\lambda_i \bar\lambda' _j)\right)}{\Delta(\Lambda)\Delta(\bar \Lambda')  }
\end{equation}
where $\Omega$ is a normalization constant. We rewrite the numerator to reintroduce $f_R$:

\begin{equation}
\label{CHIZsaddle}
\Omega\det\left(\exp(\lambda_i \bar\lambda' _j)\right)= \sum_{\vec{j}} \frac 1 {f_{\vec{j}}}  \det\left(\lambda_k^{j_i+N-i}\right)\det\left(\bar\lambda_k^{\prime j_i+N-i}\right)
\end{equation}

We have relabeled $R$ with the indices $\vec{j}$, and have rewritten the equation above accordingly. The expressions inside the determinants are monomials and correspond to the term $\prod_i \lambda_i^{j_i+N-i}+ \dots$ in $\det\left(\lambda_k^{j_i+N-i}\right)$. Thus we may expand the exponential in Eq. \eqref{CHIZsaddle} as:

\begin{equation}\label{series1}
\det\left(\exp(\lambda_i \bar\lambda' _j)\right)= \sum_{[n]} \frac 1{[ n]!} \det ((\lambda_i \bar\lambda' _j)^{n_i}) = \sum_{[n]} \frac 1{[ n]!} \det (\bar\lambda _j^{\prime n_i})\prod_i \lambda_i^{n_i} + \dots,
\end{equation}

where we have made use of the multilinearity of the determinant. The factor $[n]$ encapsulates $n_1, \dots,n_N$; then $[n]!=\prod _jn_j!$. We see that we are limited to $n_1>n_2\dots $ when we restrict ourselves to the monomials with the correct descending order; when we set $n_i = j_i+N-i$, we arrive at an explicit sum over the characters. Thus our denominator $f_{\vec{j}}$ may be computed as:

\begin{equation}
f_{\vec{j}}= \Omega^{-1} \prod_i (j_i+N-i)!,
\end{equation}

We may set $f_{\vec 0}=1$, as $\braket{0|0}=1$. Then we arrive at:

\begin{equation}
    \Omega= \prod_{i=1}^N (N-i)! 
\end{equation}
From this we can easily read off the norms of the states $\chi_R(a^\dagger)$:
\begin{equation}
    \langle \chi_{R'}(a)\chi_R(a^\dagger) \rangle= \delta_{R, R'} \frac{\prod_i (j_i+N-i)!}{\prod_{i=1}^N (N-i)!},
\end{equation}
which agrees with the well-known result of \cite{Corley:2001zk}.
\section{Symplectic and orthogonal cases}

Before repeating the analysis for the other simple lie groups, we should comment on the interpretation of the $Sp(2N)$ and $SO(N)$ theories as orientifold projections of a unitary theory. To do this, we first consider a simple toy model correponding to a single harmonic oscillator. As it turns out, this simple model captures a lot of the qualitative behaviour of the answer for symplectic and orthogonal groups. 

\subsection{A toy model for the orientifold projection}
As a warm-up, we consider a single quantum harmonic oscillator:
\begin{equation}
    [a, a^\dagger]=1.
\end{equation}
A natural basis of states for this system is the eigenstates of the occupation number operator $\hat{n}\ket{n}=n \ket{n}$. One thing that we may do with this system is to define a parity operator $\Omega= (-1)^{\hat{n}}$ and further divide the set of states into those that are mutual eigenvectors of $\hat{n}$ and $\Omega$. This gives an orthogonal decomposition of the Hilbert space of the harmonic oscillator into sectors of positive and negative parity $\mathcal{H}\cong \mathcal{H}_+ \bigoplus \mathcal{H}_{-}$, and divides all the states into even and odd states under the orientation reversal transformation
\begin{equation}
\begin{aligned}
    P: x&\rightarrow -x\\
    P: p & \rightarrow -p,
\end{aligned}
\end{equation}
where $x$ and $p$ are the position and momentum operators. Because the raising operators are monomials in $x$ and $p$, the odd parity states are created with odd numbers of raising operators and vice versa.  The operators $\frac{1}{2}\left(1\pm \Omega \right)$ respectively serve as orthogonal projection operators into $\mathcal{H}_{+}$ and $\mathcal{H}_{-}$.

What we would like to do is build coherent states in each of these two sectors of the theory. For instance, we can project a coherent state into the sector of positive parity by applying the operator $\frac{1}{2}\left(1+ \Omega \right)$:

\begin{equation}
    \frac{1}{2}\left(1+ \Omega \right)\ket{\alpha}= \frac{1}{2}\left(1+ e^{\pi i \hat{n}} \right)e^{\alpha a^\dagger}\ket{0}= \frac{1}{2}\left(e^{\alpha a^\dagger}+  e^{-\alpha a^\dagger}\right)\ket{0}= \cosh\left(\alpha a^\dagger \right)\ket{0}.
\end{equation}
We call this state $\ket{\alpha, +}$. One nice property of this state is that it is annihilated by $a^{2k+1}$ for any non-negative integer $k$. It is also an eigenstate of $a^{2}$ with eigenvalue $\alpha^2$. In this sense, we can call this a coherent state for the positive chirality sector of the model. By a similar computation, the overlap between any two of these coherent states is given by:
\begin{equation}
  \langle \beta^*,+\ket{\alpha,+}= \cosh\left( \alpha \beta\right).
\end{equation}

The case for negative parity requires more care, and will be the case that is relevant to the analysis of the $Sp(2N)$ and $SO(2N+1)$ theories. If we project a coherent state into the sector of negative chirality, we obtain the state:
\begin{equation}
      \frac{1}{2}\left(1- \Omega \right)\ket{\alpha}= \sinh\left(\alpha a^\dagger \right)\ket{0}.
\end{equation}
The issue is that this state is not a coherent state in the usual sense; when we act on the state with a lowering operator, the state won't return to the original state since the minimum ocupation number that appears in the series is $\ket{1}$. Rather, this state is also an eigenvector of $a^2$ with eigenvalue $\alpha^2$. Since the original vacuum state is annihilated by the projector $\frac{1}{2}\left(1-\Omega \right)$, the true vacuum in this sector is the state occupation number one $\ket{1}$. By a relabeling of the states for the odd sector, the coherent state can be written as
\begin{equation}
    \ket{\alpha, -}= -i\;  \text{sinc}\left( i \alpha a^\dagger \right) \ket{\tilde{0}},
\end{equation}
where $\text{sinc}(x)=\frac{\sin x }{x}$, and the new vacuum is $\ket{\tilde{0}}= \ket{1}$. A simple computation yields the norm of this coherent state:
\begin{equation}
    \langle \beta^*,- \ket{\alpha,-}= \frac{\sinh \alpha \beta}{\alpha \beta}.
\end{equation}

\subsection{The symplectic HCIZ integral}
\label{sec:symplecticHCIZs}

We now seek to expand our definition for a well-defined BPS operator averaged over the unitary group to the symplectic group: 

\begin{equation}
\label{spBPSoperator}
    F_{Sp(2N)}[\Lambda]= \frac{1}{\text{Vol}(Sp(2N))}\int_{Sp(2N)} dg \exp( \Tr(g\Lambda g^{-1} a^\dagger_Z))\ket 0,
\end{equation}
where $dg$ is the Haar measure for the symplectic group and $\text{Vol}(Sp(2N)) = \int_{Sp(2N)} dg$ is a normalization factor, which we can always rescale to one. The group elements of $Sp(2N)$ can be represented by $2N\times2N$ matrices that are both unitary and symplectic:
\begin{equation}\label{sympgroup}
\begin{aligned}
    g^\dagger g&= \mathbf{1}_{2N}\\
   g^T \Omega g&= \Omega,
\end{aligned}
\end{equation}
where $\Omega$ is a choice of anti-symmetric symmetric matrix:
\begin{equation}
    \Omega= \begin{pmatrix} 0 & \mathbf{1}_N\\
    -\mathbf{1}_N& 0
    \end{pmatrix}.
\end{equation}
The symplectic condition \eqref{sympgroup} translates into the orientifold projection of the Chan-Paton indices for the open strings ending on a stack of $2N$ $D3$ branes \cite{witten1998baryons}.
This forces the raising and lowering operators of the $Sp(2N)$ theory to satisfy the orientifold projection condition:
\begin{equation}
    \Omega \; a^\dagger_{Z}\; \Omega= (a^\dagger_Z)^T= -a^\dagger_Z,
\end{equation}
where the transpose is taken on the group indices, which we omit for clarity. This means that any operator made from traces of odd numbers of fields will automatically vanish. We choose to normalize the commutation relations for the raising and lowering operators by a factor of $\frac{1}{2}$, which will make the computation of the norm of the coherent state more transparent:
\begin{equation}
    [(a_Z)_{j}^i, (a_Z^\dagger)_k^l]= \frac{1}{2}\left(\delta_{j}^l \delta_k^i - \Omega_{jk} \Omega^{lj} \right).
\end{equation}

As with the unitary case, we wish to compute the overlap between two coherent states. This is done by applying the Campbell-Hausdorff formula; since the raising and lowering operators have different relations from the unitary case, we must check that commuting the exponentials really simplifies the norm into the form where it can be evaluated by a Harish-Chandra integral. After some algebra, we see that in the symplectic case, the exponentials can be commuted as follows:
\begin{equation}\label{hcsymp}
\begin{aligned}
    \left[\Tr\left( g a_z g^\dagger \Lambda\right), \Tr\left( h a_z^\dagger h^\dagger \bar{\Lambda}'\right)\right]&= \frac{1}{2} \Tr\left( gh \Lambda (gh)^\dagger \bar{\Lambda}'\right)+ \frac{1}{2}\Tr\left(g \Lambda g^\dagger \Omega h^T \bar{\Lambda}'^T (h^T)^{-1}\Omega \right)\\
    &= \Tr \left( gh \Lambda (gh)^\dagger \bar{\Lambda}'\right).
\end{aligned}
\end{equation}

The second term in \eqref{hcsymp} is equivalent to the first term after using the group relations \eqref{sympgroup}. This means that once again, we can compute the operator's overlap with its adjoint with the symplectic Harish-Chandra integral:

\begin{equation}
\label{eqn:spHCIZintegral}
 \bar F_{Sp(2N)}[\bar \Lambda]* F_{Sp(2N)}[\Lambda]  = \int d\tilde g \exp\left(\Tr\left( \tilde g ^{-1}\Lambda \tilde g \bar \Lambda^{\prime}\right)\right) = \mathcal{H}_{Sp(2N)}(\Lambda, \bar{\Lambda}'),
\end{equation}
where $\mathcal{H}_{Sp(2N)}(\Lambda, \bar{\Lambda}')$ is given in \cite{mcswiggen2021harish}:

\begin{equation}
\label{symplecticHCIZ}
    \mathcal{H}_{Sp(2N)}(\Lambda, \bar{\Lambda}') = \left(\prod_{p=1}^{2N-1}(2p+1)!\right)\frac{\det\left[\sinh\left(2\Lambda_j\bar{\Lambda}'_k\right)\right]^{2N}_{j,k=1}}{\Delta\left(\Lambda^{(2)}\right)\Delta\left(\bar{\Lambda}^{(2)}\right)\prod_{i=1}^{2N}\lambda_i\bar{\lambda}'_i}.
\end{equation}

 The denominator in this formula is computed using the Weyl denominator formula for the corresponding discriminant, as demonstrated in \cite{fulton2013representation, de2011separation}: 

\begin{equation}
    \boldsymbol{\Delta}_{\mathfrak{sp}(2N)}(\lambda) = \prod_j^{N} \lambda_j\prod_{1\leq j < k \leq N}\left(\lambda_j^2 - \lambda_k^2\right) = \det(\Lambda)\; \Delta\left(\Lambda^{2}\right)
\end{equation}

Thus we may rewrite Eq. \eqref{symplecticHCIZ} as:
\begin{equation}
      \boldsymbol{\Delta}_{\mathfrak{sp}(2N)}(\lambda) \boldsymbol{\Delta}_{\mathfrak{sp}(2N)}(\bar{\lambda}') \mathcal{H}_{Sp(2N)}(\Lambda,\bar{\Lambda}')= \left(\prod_{p=1}^{N-1}(2p+1)!\right) \det\left[\sinh(2\Lambda_j \bar{\Lambda}'_k) \right].
\end{equation}
The numerator can be simplified by using the identity that $\sinh(2\Lambda_j\bar{\Lambda}'_k)$ is a modified Bessel function of the first kind of order $\nu = \frac{1}{2}$, and expanding the determinant. We know that:
\begin{equation}
\begin{aligned}
    \sinh\left(2\Lambda\bar{\Lambda}'\right) = \sqrt{\pi \Lambda\bar{\Lambda}'}\;I_{\frac{1}{2}}\left(2\Lambda\bar{\Lambda}'\right) =  \sum_{m=0}^\infty\frac{2^{m+1}}{m!\left(2m+1\right)!!}\left(\Lambda\bar{\Lambda}'\right)^{2m+1}
\end{aligned}
\end{equation}

Then we can use the Cauchy-Binet formula to expand the determinant:

\begin{equation}
\begin{aligned}
    \det\left[\sinh\left(2\Lambda_i\bar{\Lambda'}_j\right)\right] = \sum_{m_i}\prod_i^{N}\frac{2^{m_i+1}}{m_i!\left(2m_i+1\right)!!}\det\left[\Lambda_j^{2m_i+1}\right]\det\left[\bar{\Lambda'}_j^{2m_i+1}\right]
\end{aligned}
\end{equation}

Thus Eq. \eqref{symplecticHCIZ} becomes:

\begin{equation}
\label{spcharacters}
    \mathcal{H}_{Sp(2N)}(\Lambda, \bar{\Lambda'}) = \sum_{m_i}\prod_i^{N}\frac{2^{m_i+1}\left(2i-1\right)!}{m_i!\left(2m_i+1\right)!!}\frac{\det\left[\Lambda_j^{2m_i}\right]\det\left[\bar{\Lambda'}_j^{2m_i}\right]}{\prod_{i<j}(\lambda_i^2-\lambda_j^2)(\bar{\lambda'}_i^2-\bar{\lambda'}_j^2)}
\end{equation}

Once again, if we set $m_i = \mu_i + N - i $, we may rewrite Eq. \eqref{spBPSoperator} as an explicit sum over the Schur polynomials:
\begin{equation}
\label{spcharacterexpansion}
   \mathcal{H}_{Sp(2N)}(\Lambda, \bar{\Lambda}') = \sum_{\mu} \frac{1}{f_{\mu}} \chi_\mu(\Lambda^2) \chi_\mu(\bar{\Lambda}'^{2}),
\end{equation}
where the coefficient in the expansion is given by
\begin{equation}
\label{spcoefficient}
    f_{\mu} = \prod_{i}^{N}\frac{\left(\mu_i + N - i \right)!\left(2\mu_i+2N-2i+1\right)!!}{2^{\mu_i+N-i+1}\left(2i-1\right)!},
\end{equation}
and the sum is taken over all integer partitions $\mu$.

This form of the expansion is natural from the point of view of the orientifold projection, since we projected out all the states with an odd number of raising operators acting on the vacuum state. Similarly, the operator that creates the coherent state must have a formal expansion of a similar form:
\begin{equation}
    \mathcal{O}_\Lambda = \int_{Sp(2N)} dg \exp\left( \Tr\left(g \Lambda g^{-1} a^\dagger_Z\right)\right)= \sum_{\mu} \frac{1}{f_\mu} \chi_\mu(\Lambda^2) \chi_\mu(\left(a^\dagger_Z\right)^2)
\end{equation}
This indicates that just as in the unitary case, the norms of states are given by the inverse of the coefficients that appear in the expansion of the Harish-Chandra integral.

\subsection{Special orthogonal groups}

\subsubsection{Odd special orthogonal group}

It is known that the Harish-Chandra integral for the odd orthogonal group is the same as that for the symplectic group. This can be thought of as a result of the $S$-duality of $\mathcal{N}=4$ super Yang-Mills theory; $S$-duality exchanges the $Sp(2N)$ and $SO(2N+1)$, while $SO(2N)$ is $S$-duality invariant \cite{witten1998baryons}. This means that the spectrum of the $Sp(2N)$ and the $SO(2N+1)$ theories are related by a change of basis. We will argue that this change of basis is simply the transpose operation on the Young diagram $\mu$ associated to a given representation.

One reason to suspect that this is the case comes from the Schur-Weyl duality for odd orthogonal and symplectic groups. It is well-known that the centralizer algebra associated to the $k$-fold tensor product of fundamental representations of $SU(N)$ is the group algebra of the symmetric group $\mathbb{C}S_k$. This means that the $k$-fold tensor product of fundamental representations of $SU(N)$ decomposes into tensor products of irreducible representations of $S_k$ and $SU(N)$:

\begin{equation}
    V_{SU(N)}^{\otimes k}\cong\bigoplus_{\lambda} \pi^{\lambda}\otimes  U_\lambda.
\end{equation}

This is more complicated for the symplectic and orthogonal groups, since the corresponding centralizer algebra is no longer a group algebra, but rather the algebra associated to the Brauer monoid. One way to understand this is that the symplectic and orthogonal lie algebras have additional invariant tensors compared to the unitary case. For tensor products of fundamental representations of unitary groups, the only invariant tensors allowed are the identity and permutation operators:
\begin{equation}
\begin{aligned}
\mathbb{I}(V_a \otimes V_b)&\rightarrow V_a \otimes V_b\\
\mathbb{P}(V_a \otimes V_b)&\rightarrow V_b \otimes V_a.
\end{aligned}
\end{equation}

Clearly these operations are invertible and generate the symmetric group $S_k$.  For orthogonal groups, there is an additional invariant tensor, called the trace operation:
\begin{equation}
    \mathbb{K}(V_a \otimes V_b)\rightarrow \mathbb{C}.
\end{equation}

These tensors are well known in the integrable spin chain literature, and are the same kind of tensors that appear in the $SO(6)$ integrable spin chain \cite{Beisert:2010jr}. Unlike the identity and permutation operators, the trace operation is not invertible, and together with the identity, it generates the Temperley-Lieb algebra $TL_k(2N)$ \cite{benkart2005tensor, ram1995characters}; the linear span of these three operations generates the Brauer algebra $B_{k}(2N)$. The importance of Brauer centralizer algebras has been emphasized in \cite{Kimura:2007wy, Ramgoolam:2008yr}, where they were used to diagonalize two-point functions in the space of gauge theory operators and their adjoints. These operators correspond to bound states of non-holomorphic giants. Brauer centralizer algebras have also been used to construct coherent states in \cite{Lin:2017vfn}.

Returning to the tensor decomposition of the $k$-fold tensor product of fundamentals of $SO(2N+1)$, the corresponding decompostition is \cite{ram1995characters}:

\begin{equation}\label{soSW}
    V_{SO(2N+1)}^{\otimes k}\cong \bigoplus_{k=0}^{\lfloor f/2 \rfloor}\bigoplus_{\lambda \vdash f-2k} D_{\lambda} \otimes V_\lambda,
\end{equation}
with $D_\lambda$ and $V_\lambda$ respectively denoting the irreducible representations of the Brauer algebra and $SO(2N+1)$. The analogous statement for the symplectic group $Sp(2N)$ exchanges $N$ with $-N$ and $V_\lambda$ with $W_{\lambda^T}$, where $W_{\lambda^T}$ is the irreducible representation of $Sp(2N)$ associated to the diagram conjugate to $\lambda$:
\begin{equation}\label{SpSW}
     V_{Sp(2N)}^{\otimes k}\cong \bigoplus_{k=0}^{\lfloor f/2 \rfloor}\bigoplus_{\lambda \vdash f-2k} D_{\lambda} \otimes W_{\lambda^T}.
\end{equation}

Since the Harish-Chandra integral involves group averages of powers of traces of the form $\Tr\left( g \Lambda g^{-1} \Lambda'\right)$, it is natural to expect that every term in expansion for the odd orthogonal groups should match to a term with the corresponding transposed Young diagram in the expansion for the symplectic integral. This might appear surprising, since the number of boxes that can appear in a column is bounded from above by $N$, while the number of boxes in a row can be arbitrary. One way of understanding this apperent mismatch is that the fundamental degrees of freedom in one description might be mapped to a bound state by S-duality. In reality, representations with arbitrary numbers of boxes in a column are possible, but will not be irreducible.

\subsubsection{Special even orthogonal group}

Extending our definition for a well-defined BPS operator to the even special orthogonal group requires a little more work. We modify the definition of $F[\Lambda]$ to reflect averaging over the even special orthogonal group:

\begin{equation}
\label{soBPSoperator}
    F_{SO(2N)}[\Lambda]= \int dO \exp( \Tr(O\Lambda O^{-1} a^\dagger_Z))\ket 0.
\end{equation}

As before, the overlap of $F[\Lambda]$ and its adjoint is the corresponding Harish-Chandra integral:

\begin{equation}
\label{eqn:soHCIZintegral}
 \bar F_{SO(2N)}[\bar \Lambda]* F_{SO(2N)}[\Lambda]  = \int d\tilde O\exp\left(\Tr\left( \tilde O ^{-1}\Lambda \tilde O \bar \Lambda^{\prime}\right)\right) = \mathcal{H}_{SO(2N)}(\Lambda, \bar{\Lambda}'),
\end{equation}
where $\mathcal{H}_{SO(2N)}(\Lambda, \bar{\Lambda}')$ is given by \cite{mcswiggen2021harish}:

\begin{equation}
\label{eqn:evensoHCIZ}
    \mathcal{H}_{SO(2N)}(\Lambda, \bar{\Lambda}') = \left(\prod_{p=1}^{N-1}(2p)!\right)\frac{\det\left[\cosh\left(2\Lambda_j\bar{\Lambda}'_k\right)\right]^{N}_{j,k=1}+\det\left[\sinh\left(2\Lambda_j\bar{\Lambda}'_k\right)\right]^{N}_{j,k=1}}{\Delta\left(\Lambda^{(2)}\right)\Delta\left(\bar{\Lambda}'^{(2)}\right)}.
\end{equation}

We note that Eq. \eqref{soBPSoperator} is invariant under an additional symmetry:

\begin{equation}
    O \rightarrow \tilde{I} O,
\end{equation}
where $\tilde{I}$ is a diagonal matrix with determinant equal to $\pm 1$. To get rid of this redundancy, we could integrate over the entire orthogonal group $O(N)$. For $SU(N)$, $Sp(2N)$ and $SO(2N+1)$, this process does not change the value of the integral. This is similar to what happens in the Kazakov-Migdal model in \cite{kazakov1993induced}, where the additional abelian part of the gauge field decouples from the collective field effective action. We also note that even though the whole integral is invariant under the parity transformation
\begin{equation}
\begin{aligned}
\tilde{P}: \Lambda&\rightarrow- \Lambda\\
\tilde{P}:\Lambda'&\rightarrow- \Lambda',
\end{aligned}
\end{equation}
the overlap is not invariant under the individual reflections of each of the eigenvalue matrices. This is because the second term is odd under transformation by individual reflections of the matrices $\Lambda$ and $\Lambda'$. Since each state must be individually invariant under this reflection, we choose to use the Harish-Chandra integral for $O(2N)$:
\begin{equation}
    \mathcal{H}_{O(2N)}=\left(\prod_{p=1}^{N-1}(2p)!\right)\frac{\det\left[\cosh\left(2\Lambda_j\bar{\Lambda}'_k\right)\right]^{N}_{j,k=1}}{\Delta\left(\Lambda^{(2)}\right)\Delta\left(\bar{\Lambda}'^{(2)}\right)}.
\end{equation}
This is precisely the matrix analogue of the norm of the coherent state for the positive parity states of a harmonic oscillator. The main difference between each of the orientifold projections is that the vacuum of each theory is charged differently under parity; the symplectic case formally begins at occupation number one of the parent theory, while the even orthogonal case begins at occupation number zero. 

We can now repeat the analysis of the previous sections with $\det\left[\cosh\left(2\Lambda_j\bar{\Lambda}'_k\right)\right]^{N}_{j,k=1}$ . We know that:

\begin{equation}
\begin{aligned}
    \cosh\left(2\Lambda\bar{\Lambda}'\right) = \sqrt{\pi\Lambda\bar{\Lambda}'}I_{-\frac{1}{2}}\left(2\Lambda\bar{\Lambda}'0\right) =  \sum_{m=0}^\infty\frac{2^m}{m!\left(2m-1\right)!!}\left(\Lambda\bar{\Lambda}'\right)^{2m}
\end{aligned}
\end{equation}

Applying the Cauchy-Binet formula yields:

\begin{equation}
    \det\left[\cosh\left(2\Lambda_i\bar{\Lambda}'_j\right)\right] = \sum_{m_i}\prod_i^{N}\frac{2^{m_i}}{m_i!\left(2m_i-1\right)!!}\det\left[\Lambda_j^{2m_i}\right]\det\left[\bar{\Lambda}_j'^{2m_i}\right]
\end{equation}

Then the Harish-Chandra integral for $O(2N)$ becomes:

\begin{equation}
\begin{aligned}
\mathcal{H}_{O(2N)}(\Lambda, \Lambda') &= \sum_{m_i}\prod_i^{N}\frac{2^{m_i}(2i-2)!}{m_i!\left(2m_i-1\right)!!}\frac{\det\left[\Lambda_j^{2m_i}\right]\det\left[\bar{\Lambda'}_j^{2m_i}\right]}{\prod_{i<j}(\lambda_i^2-\lambda_j^2)(\bar{\lambda'}_i^2-\bar{\lambda'}_j^2)}\\
\end{aligned}
\end{equation}
By setting $m_i = \mu_i + N - i$, the expression once again becomes a sum over Schur polynomials:
\begin{equation}
   \mathcal{H}_{O(2N)}(\Lambda, \Lambda')= \sum_{\mu} \frac{1}{h_\mu} \chi_\mu(\Lambda^2)\chi_\mu((\Lambda')^2),
\end{equation}
where the coefficient is now given by:
\begin{equation}
\label{socoefficient}
    h_\mu = \frac{(\mu_i +N-i)!\left(2\mu_i+2N-2i-1\right)!!}{2^{\mu_i+N-i}(2i-2)!}.
\end{equation}
Once again, we can expand the operator itself as a formal sum:
\begin{equation}
    \int_{O(N)} dO \exp\left(O \Lambda O^T a_Z^\dagger \right)= \sum_{\mu} \frac{1}{h_\mu} \chi_\mu(\Lambda^2)\chi_\mu((a_Z^\dagger)^2),
\end{equation}
which implies that the norm of the states are given by $h_\mu$.

We chose to get rid of the redundancy by integrating over $O(2N)$ rather than $SO(2N)$; in doing so, we have chosen a specific partition function. The drawback to choosing $O(2N)$ as our gauge group is that we eliminate the Pfaffian operator, which is defined as:

\begin{equation}
    Pf(\Lambda)^2 = \det(\Lambda),
\end{equation}
where $\Lambda$ is a $2n\times2n$ skew-symmetric matrix. If we make another choice and integrate over $SO(2N)$ instead, our Harish-Chandra integral becomes:

\begin{equation}
\begin{aligned}
\mathcal{H}_{SO(2N)}(\Lambda, \Lambda') &= \sum_{m_i}\prod_i^{N}\frac{2^{m_i}(2i-2)!}{m_i!\left(2m_i-1\right)!!}\frac{\det\left[\Lambda^{2m_i}\right]\det\left[\bar{\Lambda'}^{2m_i}\right]}{\prod_{i<j}(\lambda_i^2-\lambda_j^2)(\bar{\lambda'}_i^2-\bar{\lambda'}_j^2))}\\
&+ \sum_{n_i}\prod_i^{N}\frac{2^{n_i+1}\left(2i\right)!}{n_i!\left(2n_i+1\right)!!}\frac{\det\left[\Lambda^{2n_i+1}\right]\det\left[\bar{\Lambda'}^{2n_i+1}\right]}{\prod_{i<j}(\lambda_i^2-\lambda_j^2)(\bar{\lambda'}_i^2-\bar{\lambda'}_j^2))}
\end{aligned}
\end{equation}

We see that the Pfaffian of $SO(2N)$, which changes sign under a single reflection, makes an appearance in the term we previously discarded. If we write $\Lambda = X_j + iX_k$, where $X_j$ and $X_k$ are two of the six scalar fields $X_i$ in the adjoint representation of $SO(2N)$ $\mathcal{N}=4$ SYM, then $Pf(\Lambda)$ corresponds to a single BPS $D3$ brane wrapped around the non-trivial three-cycle of $\mathcal{RP}^5$ \cite{witten1998baryons, balasubramanian2002giant}. It can be considered half of a maximal giant graviton, which is identified as $\det(\Lambda)$, since the maximal giant graviton wraps around the non-trivial cycle twice.

\section{A change of basis}

One approach to diagonalizing two-point functions is to build an orthogonal basis for $Sp(2N)$ and $SO(2N)$ using local operators, as done in \cite{caputa2013basis, caputa2013operators}, which built on the restricted Schur polynomials introduced in \cite{bhattacharyya2008exact}. This is achieved by introducing a tensor $T$ in $V^{\otimes 2n}$ that has $2n$ indices and taking the sum over Wick contractions as a sum over permutations in $V^{\otimes 2n}$, or over $S_n[S_2]$. $T$ is then decomposed into irreducible components that don't mix under $S_{2n}$ when computing the two-point function. Operators are then built using projectors that commute with all of the permutations; it can be shown that these operators diagonalize the two-point function. Because these operators should be invariant in $SO(2N)$, their indices should contract in pairs. Each index corresponds to a box in the Young diagram $R$ for a tensor in representation $R$. The Young diagrams that correspond to non-zero, gauge-invariant operators have an even number of boxes in each column and row, which is to say that $2n$ is divisible by $4$, and that a square Young diagram composed of four boxes may be used a building block for the Young diagram $R$. Then the number of gauge invariant operators that can be built from $n$ fields is the number of partitions of $n/2$. We now reproduce the formula for computing two-point functions in the operator basis defined in \cite{caputa2013basis,caputa2013operators} for both $Sp(2N)$ and $SO(2N)$:

\begin{equation}
\label{caputa}
\langle\mathcal{O}_R(Z)\bar{\mathcal{O}}_S(Z)\rangle = \delta_{RS}2^n\left(\frac{d_{R/4}}{d_R}\right)^2\prod_{i \in \text{even boxes in $R$}}c_i,
\end{equation}
where $R, S$ are Young diagrams with $2n$ boxes; $R/4$ is a Young diagram with $n/2$ boxes that corresponds to the Young diagram $R$; $d_{R/4}, d_R$ are respectively the dimensions of the representations; and $c_i$ is the factor $N + a - b$ assigned to each box, where $a$ is the column index and $b$ is the row index. 

It is difficult to match our results exactly to that of \cite{caputa2013basis, caputa2013operators}, given the difference in bases. Nevertheless, we may still observe a few similarities. A natural expectation is that the HCIZ integral for a particular group has an expansion in terms of the irreducible characters of the corresponding group. An argument for this would be as follows: first we consider the exponential of the trace $\Tr\left(g \Lambda g^{-1} a_Z^\dagger \right)$. We can then expand this exponential and exchange the order of the sum or integration to evaluate the Harish-Chandra integral as in the unitary case:
\begin{equation}
  \mathcal{O}_\Lambda= \sum_{m=0}^\infty \frac{1}{m!}\int dg\Tr\left(g \Lambda g^{-1} a_Z^\dagger \right)^m.
\end{equation}
We may try to express each term as a character of the corresponding group by taking traces of Eq. \eqref{soSW} or Eq. \eqref{SpSW}. Formally, this gives an expansion for the Harish-Chandra integral as a sum of infinitesimal characters evaluated on the Lie algebra. For $U(N)$, this is not a problem, because the formulas for Schur polynomials make sense when evaluated on the Lie algebra. This does not seem to be the case for $Sp(2N)$ and $SO(N)$. Even then, one may try to make sense of this formal expansion in order to get a formula for the coefficients. If one extrapolates the answer for the unitary case, the expectation would be that the coefficients are ratios of dimensions of irreducible representatiations of the group and the corresponding centralizer algebra. This turns out to be partially true, since the coefficient associated to single row representations in Eq. \eqref{spcharacterexpansion} seems to agree precisely with the ratio of
\begin{equation}
    c_\mu = \frac{2^m d^\mu}{m! D_\mu},
\end{equation}
where $\mu$ is a partition of $m$, $d^\mu$ is the dimension of the irreducible representation of the symmetric group $S_m$, and $D_\mu$ is the dimension of the corresponding symplectic group representation. This is clearly different from Eq. \eqref{caputa}, but we note that the number of partitions of $n/2$ is the dimension of $S_{n/2}$, which is equivalent to $d_{R/4}$. Thus we have preserved the characteristic of the coefficient as a function of the ratio of the dimension of the irrep of the corresponding symmetric group to the dimension of the gauge group representation.

We now make the observation that since we perform the character expansion using Schur polynomials, which present as ratios of determinants, our basis is directly linked to free fermions; after all, the Schur functions correspond to free fermion wave functions \cite{Corley:2001zk, berenstein2004toy}. We return to the results of \cite{caputa2013operators}, where it is shown that the character of the local operator can be written in terms of a Schur polynomial of the matrix of the operator's eigenvalues. Thus the character of the operator has the interpretation of the Slater determinant of $N/2$ single particle wave functions, or $N/2$ fermions moving in an external harmonic oscillator potential. So we may conclude that our basis describes the same dynamics as the operator basis constructed in \cite{caputa2013basis, caputa2013operators}.

\section{Discussion}

In this paper, we extended the method of computing the norms of half BPS coherent states through localization \cite{Berenstein:2022srd} to theories with the gauge groups $Sp(2N)$, $SO(2N+1)$, and $SO(2N)$. We did this by constructing coherent states averaged over a group orbit from each group and computing the norm of these states through the symplectic and special orthogonal Harish-Chandra integrals. The integration over the group may be viewed as a sort of path integral over the emergent world-volume gauge symmetry of a stack of $N$ giant gravitons inside $AdS_5 \times \mathbb{RP}^5$; the norm of the state gives the effective action of this theory. Curiously enough, these types of integrals first appeared in models of induced QCD. By expanding the Harish-Chandra integrals, we found that each integral admits an expression as a sum of unitary characters. This matches what one would expect of an orientifold projection of a $U(2N)$ gauge theory; all the states that are spanned by the coherent states are "doubled" versions of those in the original theory. In particular, the coherent states considered here do not span the complete spectrum of the free $Sp(2N)$ and $SO(2N)$ theories. This is because the Harish-Chandra integral is only able to capture information from tensor contractions of the invariant tensors of the unitary group (meaning all products of traces). It is likely that some of the data corresponding to worldsheets with cross-caps is missing.

As in the unitary case, the coefficient associated with the characters in this series expansion computes the overlap of the corresponding Schur polynomials of the operators $(a)^i_j$ and $(a^\dagger)^i_j$. Our method should be contrasted to other constructions of basis of operators for the $Sp(2N)$ and $SO(2N)$ theories \cite{caputa2013basis, caputa2013operators}, since our construction uses group theoretic objects more closely associated to each group. We conclude with some comments and an outline of open questions.

\subsection*{Connection to symplectic and orthogonal characters}

A natural question is to ask is why Schur polynomials appear in the expansion for the symplectic Harish-Chandra integral, as opposed to sympletic Schur polynomials. If one tries to evaluate the symplectic Schur polynomials on a Cartan element of the Lie algebra in the most na\"ive way,
\begin{equation}
\label{spschurpoly}
    sp_\lambda (X)= \frac{\det \left[ x_j^{\lambda_i-i+1}- \Bar{x}_1^{\lambda_i-i+1}\right]}{\det \left[ x_j^{N-j+1}- \Bar{x}_1^{N-j+1}\right]},
\end{equation}
by replacing $\bar{x}_i$ with $-x_i$ instead of $1/x_i$, one obtains a suggestive formula:
\begin{equation}
    sp_{\lambda}(X) sp_\lambda (Y)\sim \frac{\det \left[x_a^{\lambda_b+N-b+1}\right]}{\Delta(x_i^2) \prod_{c=1}^N x_c}\frac{\det \left[y_a^{\lambda_b+N-b+1}\right]}{\Delta(y_i^2) \prod_{c=1}^N y_c}.
\end{equation}
This can be recognized as the terms in the expansion for the function:
\begin{equation}
   \frac{\det \left(\sinh(x_i y_j)\right)}{ \boldsymbol{\Delta}_{\mathfrak{sp(n)}}(x)  \boldsymbol{\Delta}_{\mathfrak{sp(n)}}(y)}\sim \mathcal{H}_{Sp(2N)}(x,y).
\end{equation} The main difficulty with making this a precise equality comes from the fact that the denominator and numerator of Eq. \eqref{spschurpoly} have zeros that need to cancel between each other, leaving an ambiguity for the normalization of the symplectic characters. Another issue is that different choices of representations appear to lead to the same polynomial. This is expected, since irreducible representations can appear with multiplicities in the decompostion of tensor products. However, by adding information from the centralizer algebra, one should be able to differentiate between irreducible representations. This additional data is precisely the $1/N$ corrections coming from cross-caps. This idea seems to suggest that there might be a refined version of the Harish-Chandra integral that takes into account the contributions from cross-cap states that are missing in the original integral. This would give an explicit connection between the representation theory of the  Weyl group of $Sp(2N)$ \cite{fulton2013representation}, and the Brauer algebra \cite{ram1995characters}.

Another connection between the symplectic Harish-Chandra integral and the symplectic characters comes from their generalizations to continuous Schur polynomials. A similar $\sinh[\lambda_jx_i]$ term makes an appearance in the continuous symplectic Schur function, which is defined in \cite{bisi2019point} as:

\begin{equation}
\label{eqn:continuousschur}
\begin{aligned}
sp_{\Lambda}^{\text{cont}}\left(X\right) &= \frac{\det\left[\sinh\left(\lambda_jx_i\right)\right]}{\prod_{1\leq i < j \leq N}\left(\lambda_i^2 - \lambda_j^2\right)\prod_{i=1}^N\left(\lambda_i\right)}.
\end{aligned}
\end{equation} 

Notice that up to a factor of the discriminant of $x_i$, the continuous Schur functions agree with the symplectic Harish-Chandra integral. The continuous Schur function may then be written in the form of Eq. \eqref{spcharacterexpansion}, where the determinant is folded into the coefficient $f_\mu$. The presence of the Harish-Chandra integral implies that localization occurs in this calculation. An important point is that the continuous Schur functions are defined by a diferent integral formula in \cite{bisi2019point}:

\begin{equation}
    sp^{cont}_{\Lambda}\left(X\right) = \int_{GT_{2N}\left(X \right)}  \prod_{k=1}^{N}e^{\lambda_k\left(2|z_{2k-1}|-|z_{2k-2}|-|z_{2k}|\right)}dz_{i,j},
\end{equation}
where $GT_{2N}(\Lambda)$ is the set of all continuous Gelfand-Tsetlin patterns of shape $\Lambda$. Roughly, speaking, an integer point $\mu$ in this space can be associated to a Young diagram  $\mu$. The fact that this integral evaluates to what appears to be a sum of over integer points (Young diagrams) seems to suggest that there is some sort of localization in this space. Curiously enough, this integral is somewhat remniscient of a momentum space amplitude. For instance, taking $\lambda$ to be big with $\lambda x$ held fixed, the value of the integral divided by the appropriate discriminant remains fixed, but the integration region shrinks to points where 
\begin{equation}
    2|z_{2k-1}|-|z_{2k-2}|-|z_{2k}|=0.
\end{equation} Each of these points should correspond to a particular symplectic Schur polynomial .
This is somewhat suggestive of some sort of worldsheet localization for a tensionless string \cite{Gopakumar:2005fx, Razamat:2008zr}, where the integral over the worldsheet moduli space is expected to localize to a certain set of integer points.

\subsection*{Connection to quantum Schubert calculus}

It is well known that the Schur symmetric functions are related to Schubert classes, which form an integer basis for the cohomology ring of the Grassmannian. The product of two Schubert classes may be expanded as a a linear combination of Schubert classes summed over the given partitions $\nu$ \cite{postnikov2005affine}:

\begin{equation}
\sigma_\lambda \cdot \sigma_\mu = \sum_{\nu}c^\nu_{\lambda\mu}\sigma_\nu.
\end{equation}

The different Schubert classes are represented by $\sigma_{\mu},\sigma_{\nu},\sigma_{\lambda} $ and $c^\nu_{\lambda\mu}$ represent the Littlewood-Richardson coefficients. A well-known result is that every Schubert class can be associated with a Schur polynomial; the connection is seen by noting that the cohomology product mirrors the way the product of two Schur functions can be expanded as a linear combination of ordinary Schur functions \cite{macdonald1998symmetric}:

\begin{equation}
s_\lambda \cdot s_\mu = \sum_{\nu}c^\nu_{\lambda\mu}s_\nu.
\end{equation}
This is directly related to the three point function of coherent states for the $U(N)$ theory. This is because after applying the Campbell-Hausdorff formula, one obtains an integral over a complex Grassmanian. 

It should be noted that the skew Schur polynomial can be expanded in the same basis in a similar way \cite{macdonald1998symmetric}:

\begin{equation}
s_{\lambda/\mu} = \sum_{T \in SSYT\left(\lambda/\mu\right)}\bf{x}^T = \sum_\nu c^\lambda_{\mu\nu}s_\nu,
\end{equation}

where $SSYT$ refers to the skew Schur Young tableaux. We note that the even symplectic Schur functions we touch upon in the previous subsection can be written as a sum of skew Schur functions summed over Frobenius coordinates.

We are interested in computing the Gromov-Witten invariants, which appear in the \textit{quantum} product of two Schubert classes, which is defined on the small quantum cohomology ring of the Grassmannian, $QH^*\left(Gr_{kn}\right)$. $QH^*\left(Gr_{kn}\right)$ is defined as the tensor product of the cohomology ring of the Grassmannian and the polynomial ring $\mathbb{Z}\left[q\right]$, where $q$ is a variable of degree $n$. The quantum product of two Schubert classes, then, is defined in \cite{postnikov2005affine} as:

\begin{equation}
    \sigma_\lambda*\sigma_\mu = \sum_{d,\nu}q^d C^{\nu,d}_{\lambda\mu}\sigma_\nu,
\end{equation}

where $d$ is a non-negative integer such that $|\nu| = |\lambda| + |\mu| - dn$, and $C^{\nu,d}_{\lambda\mu}$ are the Gromov-Witten invariants. Toric Schur functions are defined in \cite{postnikov2005affine} to correspond to cylindric diagrams of shape $\lambda[r]/\mu[s]$, which are defined as finite subsets of $\mathcal{C}_{nk} = \mathbb{Z}^2/(-k, n-k)\mathbb{Z}$. We label these toric Schur functions with the shape $\lambda/d/\mu$, where $d = r-s$. Since skew Schur functions are toric Schur functions when $d = 0$, it should not come as a surprise that toric Schur functions may be expanded in the Schur basis just as the former are in \cite{postnikov2005affine}:

\begin{equation}
    s_{\lambda/d/\mu} = \sum_{\nu}C^{\lambda, d}_{\mu\nu}s_\nu
\end{equation}

The main difference is that the Gromov-Witten invariants have replaced the classical Littlewood-Richardson coefficients. This replacement should correspond to replacing a centralizer algebra by a Hecke algebra. Thus once again, it should not come as a surprise that the Gromov-Witten invariants can be given as an alternating sum of classical Littlewood-Richardson coefficients, as demonstrated in \cite{BERTRAM1999728}. The problem with this approach is that there are too many Littlewood-Richardson coefficients to keep track of, which makes the computation unwieldy. It is known that Gromov-Witten invariants in other contexts may be computed through localization \cite{liu2013localization, bertiger2020equivariant}. It would be interesting if the Harish-Chandra integral can be extended to the toric Schur functions to obtain a combinatorial formula for the Gromov-Witten invariants. This may also shed light on the problem of computing the normalization of three-point functions, where the Gromov-Witten invariants appear as coefficients. 

\acknowledgments

We would like to thank D. Berenstein for helpful discussions. SW's research was supported in part by the Department of Energy under grant DE-SC0019139.






\bibliographystyle{JHEP}
	\cleardoublepage

\renewcommand*{\bibname}{References}

\bibliography{references}

\providecommand{\href}[2]{#2}\begingroup\raggedright\begin{thebibliography}{10}

\bibitem{Berenstein:2022srd}
D.~Berenstein and S.~Wang, {\it {BPS coherent states and localization}},  {\em
  JHEP} {\bf 08} (2022) 164, [\href{http://arxiv.org/abs/2203.15820}{{\tt
  arXiv:2203.15820}}].

\bibitem{Jiang:2019xdz}
Y.~Jiang, S.~Komatsu, and E.~Vescovi, {\it {Structure constants in $
  \mathcal{N} $ = 4 SYM at finite coupling as worldsheet g-function}},  {\em
  JHEP} {\bf 07} (2020), no.~07 037,
  [\href{http://arxiv.org/abs/1906.07733}{{\tt arXiv:1906.07733}}].

\bibitem{Chen:2019gsb}
G.~Chen, R.~de~Mello~Koch, M.~Kim, and H.~J.~R. Van~Zyl, {\it {Absorption of
  closed strings by giant gravitons}},  {\em JHEP} {\bf 10} (2019) 133,
  [\href{http://arxiv.org/abs/1908.03553}{{\tt arXiv:1908.03553}}].

\bibitem{Budzik:2021fyh}
K.~Budzik and D.~Gaiotto, {\it {Giant gravitons in twisted holography}},
  \href{http://arxiv.org/abs/2106.14859}{{\tt arXiv:2106.14859}}.

\bibitem{Gaiotto:2021xce}
D.~Gaiotto and J.~H. Lee, {\it {The Giant Graviton Expansion}},
  \href{http://arxiv.org/abs/2109.02545}{{\tt arXiv:2109.02545}}.

\bibitem{Murthy:2022ien}
S.~Murthy, {\it {Unitary matrix models, free fermion ensembles, and the giant
  graviton expansion}},  \href{http://arxiv.org/abs/2202.06897}{{\tt
  arXiv:2202.06897}}.

\bibitem{Corley:2001zk}
S.~Corley, A.~Jevicki, and S.~Ramgoolam, {\it {Exact correlators of giant
  gravitons from dual N=4 SYM theory}},  {\em Adv. Theor. Math. Phys.} {\bf 5}
  (2002) 809--839, [\href{http://arxiv.org/abs/hep-th/0111222}{{\tt
  hep-th/0111222}}].

\bibitem{Corley:2002mj}
S.~Corley and S.~Ramgoolam, {\it {Finite factorization equations and sum rules
  for BPS correlators in N=4 SYM theory}},  {\em Nucl. Phys. B} {\bf 641}
  (2002) 131--187, [\href{http://arxiv.org/abs/hep-th/0205221}{{\tt
  hep-th/0205221}}].

\bibitem{brown2008diagonal}
T.~W. Brown, P.~Heslop, and S.~Ramgoolam, {\it Diagonal multi-matrix
  correlators and bps operators in n= 4 sym},  {\em Journal of High Energy
  Physics} {\bf 2008} (2008), no.~02 030.

\bibitem{bhattacharyya2008exact}
R.~Bhattacharyya, S.~Collins, and R.~de~Mello~Koch, {\it Exact multi-matrix
  correlators},  {\em Journal of High Energy Physics} {\bf 2008} (2008), no.~03
  044.

\bibitem{brown2009diagonal}
T.~W. Brown, P.~Heslop, and S.~Ramgoolam, {\it Diagonal free field matrix
  correlators, global symmetries and giant gravitons},  {\em Journal of High
  Energy Physics} {\bf 2009} (2009), no.~04 089.

\bibitem{kimura2012correlation}
Y.~Kimura, {\it Correlation functions and representation bases in free n= 4
  super yang--mills},  {\em Nuclear Physics B} {\bf 865} (2012), no.~3
  568--594.

\bibitem{caputa2010spectral}
P.~Caputa, C.~Kristjansen, and K.~Zoubos, {\it On the spectral problem of
  $\mathcal{N}= 4$ sym with orthogonal or symplectic gauge group},  {\em
  Journal of High Energy Physics} {\bf 2010} (2010), no.~10 1--22.

\bibitem{caputa2013basis}
P.~Caputa, R.~de~Mello~Koch, and P.~Diaz, {\it A basis for large operators in
  n= 4 sym with orthogonal gauge group},  {\em Journal of High Energy Physics}
  {\bf 2013} (2013), no.~3 1--28.

\bibitem{caputa2013operators}
P.~Caputa, R.~de~Mello~Koch, and P.~Diaz, {\it Operators, correlators and free
  fermions for so (n) and sp (n)},  {\em Journal of High Energy Physics} {\bf
  2013} (2013), no.~6 1--39.

\bibitem{witten1998baryons}
E.~Witten, {\it Baryons and branes in anti de sitter space},  {\em Journal of
  High Energy Physics} {\bf 1998} (1998), no.~07 006.

\bibitem{berenstein2004toy}
D.~Berenstein, {\it A toy model for the ads/cft correspondence},  {\em Journal
  of High Energy Physics} {\bf 2004} (2004), no.~07 018.

\bibitem{morozov2010unitary}
A.~Y. Morozov, {\it Unitary integrals and related matrix models},  {\em
  Theoretical and Mathematical Physics} {\bf 162} (2010), no.~1 1--33.

\bibitem{mcswiggen2021harish}
C.~McSwiggen, {\it The harish-chandra integral: An introduction with examples},
   {\em L’Enseignement Math{\'e}matique} {\bf 67} (2021), no.~3 229--299.

\bibitem{fulton2013representation}
W.~Fulton and J.~Harris, {\em Representation theory: a first course}, vol.~129.
\newblock Springer Science \& Business Media, 2013.

\bibitem{de2011separation}
J.~de~Gier and A.~Ponsaing, {\it Separation of variables for symplectic
  characters},  {\em Letters in Mathematical Physics} {\bf 97} (2011), no.~1
  61--83.

\bibitem{Beisert:2010jr}
N.~Beisert et~al., {\it {Review of AdS/CFT Integrability: An Overview}},  {\em
  Lett. Math. Phys.} {\bf 99} (2012) 3--32,
  [\href{http://arxiv.org/abs/1012.3982}{{\tt arXiv:1012.3982}}].

\bibitem{benkart2005tensor}
G.~Benkart and D.~Moon, {\it Tensor product representations of temperley-lieb
  algebras and chebyshev polynomials},  {\em Representations of Algebras and
  Related Topics, in: Fields Inst. Commun., Amer. Math. Soc., Providence, RI}
  {\bf 45} (2005) 57--80.

\bibitem{ram1995characters}
A.~Ram, {\it Characters of brauer’s centralizer algebras},  {\em Pacific
  journal of Mathematics} {\bf 169} (1995), no.~1 173--200.

\bibitem{Kimura:2007wy}
Y.~Kimura and S.~Ramgoolam, {\it {Branes, anti-branes and brauer algebras in
  gauge-gravity duality}},  {\em JHEP} {\bf 11} (2007) 078,
  [\href{http://arxiv.org/abs/0709.2158}{{\tt arXiv:0709.2158}}].

\bibitem{Ramgoolam:2008yr}
S.~Ramgoolam, {\it {Schur-Weyl duality as an instrument of Gauge-String
  duality}},  {\em AIP Conf. Proc.} {\bf 1031} (2008), no.~1 255--265,
  [\href{http://arxiv.org/abs/0804.2764}{{\tt arXiv:0804.2764}}].

\bibitem{Lin:2017vfn}
H.~Lin and K.~Zeng, {\it {A construction of quarter BPS coherent states and
  Brauer algebras}},  {\em Adv. Theor. Math. Phys.} {\bf 24} (2020), no.~5
  1111--1169, [\href{http://arxiv.org/abs/1709.10093}{{\tt arXiv:1709.10093}}].

\bibitem{kazakov1993induced}
V.~Kazakov and A.~Migdal, {\it Induced gauge theory at large n},  {\em Nuclear
  Physics B} {\bf 397} (1993), no.~1-2 214--238.

\bibitem{balasubramanian2002giant}
V.~Balasubramanian, M.~Berkooz, A.~Naqvi, and M.~J. Strassler, {\it Giant
  gravitons in conformal field theory},  {\em Journal of High Energy Physics}
  {\bf 2002} (2002), no.~04 034.

\bibitem{bisi2019point}
E.~Bisi and N.~Zygouras, {\it Point-to-line polymers and orthogonal whittaker
  functions},  {\em Transactions of the American Mathematical Society} {\bf
  371} (2019), no.~12 8339--8379.

\bibitem{Gopakumar:2005fx}
R.~Gopakumar, {\it {From free fields to AdS: III}},  {\em Phys. Rev. D} {\bf
  72} (2005) 066008, [\href{http://arxiv.org/abs/hep-th/0504229}{{\tt
  hep-th/0504229}}].

\bibitem{Razamat:2008zr}
S.~S. Razamat, {\it {On a worldsheet dual of the Gaussian matrix model}},  {\em
  JHEP} {\bf 07} (2008) 026, [\href{http://arxiv.org/abs/0803.2681}{{\tt
  arXiv:0803.2681}}].

\bibitem{postnikov2005affine}
A.~Postnikov, {\it Affine approach to quantum schubert calculus},  {\em Duke
  Mathematical Journal} {\bf 128} (2005), no.~3 473--510.

\bibitem{macdonald1998symmetric}
I.~G. Macdonald, {\em Symmetric functions and Hall polynomials}.
\newblock Oxford university press, 1998.

\bibitem{BERTRAM1999728}
A.~Bertram, I.~Ciocan-Fontanine, and W.~Fulton, {\it Quantum multiplication of
  schur polynomials},  {\em Journal of Algebra} {\bf 219} (1999), no.~2
  728--746.

\bibitem{liu2013localization}
C.~Liu, {\it Localization in gromov-witten theory and orbifold gromov-witten
  theory, handbook of moduli, volume ii, 353425, adv. lect. math.,(alm) 25},
  2013.

\bibitem{bertiger2020equivariant}
A.~Bertiger, D.~Ehrlich, E.~Mili{\'c}evi{\'c}, and K.~Taipale, {\it An
  equivariant quantum pieri rule for the grassmannian on cylindric shapes},
  {\em arXiv preprint arXiv:2010.15395} (2020).

\end{thebibliography}\endgroup
\end{document}